\documentclass[prl,twocolumn,nofootinbib,amsmath,amsfonts,amssymb]{revtex4}

\usepackage{graphicx,bm,color}

\usepackage{natbib}
\newcommand{\be}{\begin{equation}}
\newcommand{\ee}{\end{equation}}
\newcommand{\bea}{\begin{eqnarray}}
\newcommand{\eea}{\end{eqnarray}}

\newcommand{\ds}{{\sf DarkSUSY}}

\begin{document}

\title{Antiproton and Radio Constraints on the Dark Matter Interpretation\\ of the Fermi Gamma Ray  Observations of the Galactic Center}

\author{Torsten Bringmann}
\affiliation{~II. Institute for Theoretical Physics, University of Hamburg, Luruper Chaussee 149, D-22761 Hamburg, Germany}
\email{torsten.bringmann@desy.de}

\date{3 Nov 2009}

\begin{abstract}
Recently, it was suggested that the gamma rays observed by the Fermi Gamma Ray Space Telescope from the direction of the galactic center could surprisingly well be described by a dark matter annihilation scenario, both in terms of their angular distribution and energy spectrum. Here, I point out that such a scenario would be in considerable tension with existing antiproton data, in particular the new PAMELA measurements. Radio data are even more constraining, thus disfavoring the dark matter hypothesis and making an astrophysical explanation of the observations much more plausible. 

\end{abstract}


\maketitle

The innermost region of the Milky Way is one of the classical targets for indirect dark matter (DM) searches \cite{dmgcgeneral}. In fact, the expected high DM density close to the galactic center would make it the single brightest DM annihilation source in the sky. This apparently favorable situation for the hunt for DM signals is, however, obscured by the fact that the region is also astrophysically very rich and complex; as a consequence, it will generally be difficult to unambiguously distentangle a DM signal from the resulting, still not very well understood background \cite{Zaharijas:2006qb}.

In the past, the potential observation of a DM signal from the galactic center region has already been claimed a couple of times -- in particular in gamma rays \cite{gcgamma}, but also in microwaves \cite{haze} or the annihilation radiation from positrons \cite{integral}. While there still might remain a bit of controversy in some of these cases, evidence is certainly not compelling and the general picture has been that more refined analyses and new data tended to disfavor the DM hypotheses previously put forward \cite{nodm} (see, however, \cite{Dobler:2009xz}). The most recent in this series of claims is the 'possible evidence for dark matter annihilation' seen by Goodenough  \& Hooper \cite{Goodenough:2009gk}. They find that the gamma rays from this direction, as observed by the Fermi Gamma Ray Space Telescope \cite{fermi}, can well be described by a DM annihilation scenario where the DM particle has a mass of $m_\chi\approx28\,$GeV and annihilates purely into $b\bar b$ final states, with a cross section of $\langle\sigma v\rangle\approx9\cdot10^{-26}{\rm cm}^3{\rm s}^{-1}$. For the dark matter distribution in the halo, a profile with $\rho_\chi(r)\propto r^{-1.1}$ has to be invoked, i.e.~somewhat steeper than the standard NFW profile \cite{NFW}. Rather than arguing for or against the plausibility of this scenario, and whether or not potential astrophysical backgrounds were adequately taken into account, let us here check the internal consistency of this proposal.

\begin{figure}[t]
  \includegraphics[width=\columnwidth]{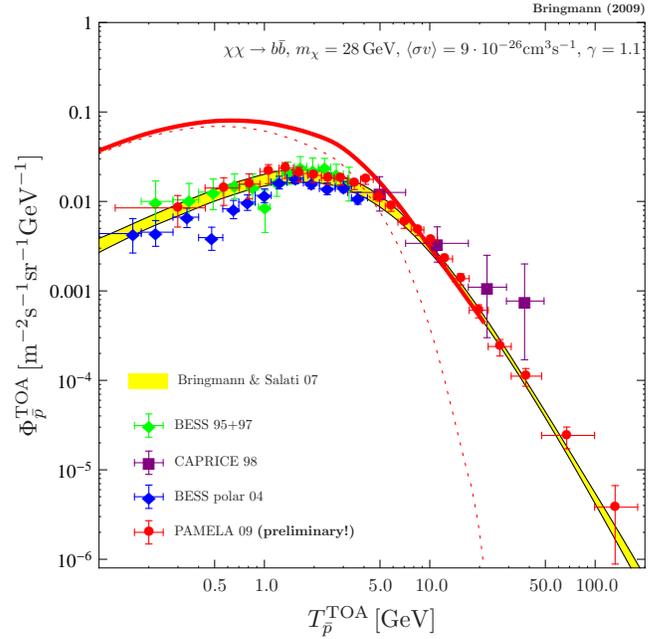}
  \caption{This Figure shows the antiproton flux as measured by BESS '95/'97 \cite{bess9597}, CAPRICE \cite{caprice}, BESS polar \cite{besspolar} and, most recently, PAMELA \cite{pam}. The yellow band shows the predicted flux in secondary antiprotons, taking into account the full range of uncertainty in the propagation parameters \cite{Bringmann:2006im}. The dotted line indicates the primary antiprotons produced in the DM model by Goodenough \& Hooper \cite{Goodenough:2009gk} and the thick red line shows the resulting total flux as calculated with \ds\ \cite{ds}, assuming the smallest allowed secondary flux.}
    \label{pbarflux}
\end{figure}

Apart from gamma rays, antiprotons are also an interesting means of indirect DM detection as they are inevitably produced whenever it is kinematically possible and the annihilation products have color charge \cite{pbar}. Unlike gamma-rays, however, they do not travel unperturbed through the galaxy but are deflected by inhomogeneities in the galactic magnetic fields. While this introduces a certain amount of uncertainty, the main aspects of antiproton propagation are well understood and can nicely be described in terms of semi-analytic diffusion models where the parameters are adjusted to fit existing cosmic ray data, in particular the observed boron over carbon ratio $B/C$ \cite{Donato:2003xg}. As a result, the expected background flux of \emph{secondary} antiprotons, mainly produced in the collisions of cosmic ray protons with interstellar matter,  is tightly constrained, even well beyond the currently accessible energies. For illustration, Fig.~\ref{pbarflux} shows the expectation for this background flux, fully taking into account the uncertainty in the propagation parameters, as predicted roughly three years ago \cite{Bringmann:2006im}. The remarkable agreement with the subsequent data of BESS polar \cite{besspolar} and (still preliminary) PAMELA \cite{pam} provides an important test for the underlying diffusion model.

In the same Figure, the total antiproton flux in the DM model proposed by Goodenough \& Hooper is indicated, as calculated by using \ds\ \cite{ds} with its default set of propagation parameters (implementing the procedure described in \cite{Bergstrom:1999jc}). As it stands, the resulting antiproton excess at energies below about 5 GeV would clearly not be compatible with the data. One has to keep in mind, however, that the $B/C$ analysis of the allowed range of propagation parameters is limited to sources in the disk; most of the annihilations, on the other hand, take place in the halo, so one cannot expect the predictions for the
flux of \emph{primary} antiprotons (produced directly by DM annihilations) to be as tightly constrained as for secondaries. By adopting extreme sets of propagation parameters, still compatible with $B/C$, the flux of primary antiprotons could, in fact, be up to an order of magnitude smaller (or larger) than what is shown in Fig.~\ref{pbarflux}  \cite{Donato:2003xg}.

\begin{figure}[t]
  \includegraphics[width=\columnwidth]{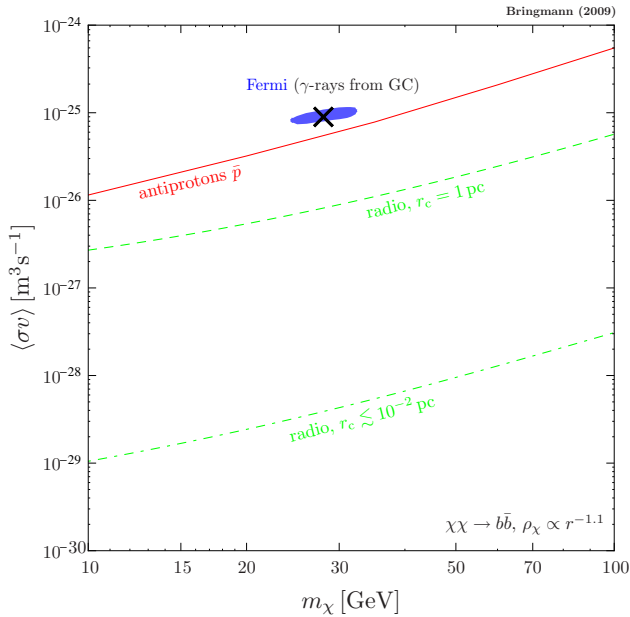}
  \caption{The shaded region indicates, roughly, the combination of DM mass $m_\chi$ and annihilation cross section $\langle\sigma v\rangle$ that could in principle describe the gamma ray spectrum observed by Fermi from the direction of the galactic center; the cross corresponds to the best fit model obtained in \cite{Goodenough:2009gk} which was also used in Fig.~\ref{pbarflux}. Models above the solid line are excluded due to the overproduction of antiprotons; the dashed (dash-dotted) line shows the constraints from radio observations, assuming a cutoff in the DM density profile at $r_{\rm c}=1\,$pc ($r_{\rm c}\lesssim10^{-2}\,$pc).}
    \label{radiofig}
\end{figure}

 Including the new (PAMELA, in particular) data sets in the analysis, on the other hand, would constrain the allowed range of propagation parameters even more \cite{DiBernardo:2009ku}. 
For a given annihilation cross section, the annihilation signal enhancement that is expected due to the effect of DM clumps in the halo (often referred to as "boost factor") is, furthermore, probably stronger for antiprotons than for gamma rays, thus making the antiproton bounds somewhat more constraining than discussed so far. The reason for this is that for a spiky DM profile like the one considered here, the largest contribution to the total flux in gamma rays comes from the innermost part of the galaxy, where clumps most likely have been completely disrupted by the strong tidal forces; antiprotons probe a much larger portion of the halo and the total flux is not very sensitive to the region close the center. Taking into account these considerations, I will in the following require that  the primary flux from DM annihilations alone should not exceed five times the maximally expected flux of background (i.e. secondary) antiprotons; while a more detailed analysis of the uncertainties involved would certainly be interesting, it is clearly beyond the scope of the present letter. The resulting constraint in the $\langle\sigma v\rangle$ vs. $m_\chi$ plane is shown in Fig.~\ref{radiofig} and falls well below the region that could describe the Fermi gamma ray data in a DM setup like the one proposed in \cite{Goodenough:2009gk}.

By changing the propagation model, or adopting extreme propagation parameters that are barely excluded by the $B/C$ analysis, one could probably still argue for the possibility to evade the antiproton constraints presented in Fig.~\ref{radiofig}.
However, even more powerful constraints can be derived by considering electrons and positrons as DM annihilation products and the synchroton radiation they emit when propagating through the galactic magnetic field.  For the galactic center, it has earlier been pointed out that this usually provides more stringent bounds than for gamma rays, or photons at other wavelengths, especially for low mass DM models \cite{radiogc}. 
In the monochromatic approximation for the synchroton emission the resulting luminosity from a cone towards the galactic center can be written as \cite{radioproc}
\be
  \label{lum}
   \nu\,L(\nu)=2\pi\frac{\langle\sigma v\rangle}{m_{\chi^2}}\int dr\,r^2{\rho_\chi}^2(r)\,E_{e^\pm}N(E_{e^\pm})\,.
\ee
In this expression, $E_{e^\pm}=\sqrt{4\pi m_e^3\nu/B}=0.43\,{\rm GeV}(\nu/{\rm GHz})^\frac{1}{2}(B/{\rm mG})^{-\frac{1}{2}}$ is the energy of an emitted electron or positron at which synchroton radiation of frequency $\nu$ is emitted and $N(E)$ is the number of electrons and positrons, per annihilation, above that energy. The strongest constraints now derive from comparing the above luminosity to radio data at the lowest observed frequency, $\nu=0.408\,$GHz  \cite{davis} (the observed region corresponds to a distance of $\sim0.14\,$pc from the galactic center). 

Before being able to derive these bounds, one needs of course also a model for the magnetic field in the innermost galaxy.
As an upper bound on its strength, one may consider it to be in equipartition up to the accretion radius $R_{\rm acc}\approx0.04\,$pc of the black hole, with magnetic pressure and kinetic energy of the infalling matter  balancing each other, which leads to \mbox{$B(r<R_{\rm acc})\approx7.2\, {\rm mG}\,(0.04\,{\rm pc}/r)^{5/4}$} \cite{Aloisio:2004hy}; above this, conservation of the magnetic flux dictates a scaling \mbox{$B(r)\propto r^{-2}$}, which in the following will be assumed down to the typical, constant galactic value of $B\sim\mu$G. As the opposite extreme of a rather small magnetic field, one may assume it to stay constant below $R_{\rm acc}$.  

Following thus the same procedure as in \cite{radioproc}, 
the resulting constraints on the DM mass and annihilation cross section are also included in Fig.~\ref{radiofig}.
Here, an artificial cutoff of the DM profile was introduced at $r_{\rm c}=1\,$pc, below which the DM density was assumed to stay constant. This is an extremely conservative assumption since such a cutoff would usually only be expected to be set by the annihilation rate \cite{Berezinsky:1992mx}, at a scale several orders of magnitude smaller than this. As also indicated in the Figure,  a smaller cutoff would further strengthen the radio constraints by a factor of up to about 200 (for $r_{\rm c}\lesssim10^{-2}\,$pc, these constraints actually do no longer depend on the cutoff for the halo profile considered here). 

It should also be stressed that the radio constraints shown in Fig.~\ref{radiofig} are only very weakly dependent on the assumptions about the magnetic field  -- in contrast to the assumptions about the DM profile which enters quadratically in Eq.~(\ref{lum}) -- since the main contribution comes from not too small radial distances, where the uncertainty in the magnetic field is not very big; varying the magnetic field configuration between the two extreme cases mentioned before leads to barely visible differences in the bounds. Note also, from Eq.~(\ref{lum}), that the luminosity actually \emph{increases} for smaller values of $B$, so the (somewhat counterintuitive) assumption of a more a less constant magnetic field at the $\mu$G level extending to the very central part of the galaxy would lead to even \emph{stronger} bounds.

At this point, one may wonder whether there is a chance to modify the proposed characteristics of the DM model in such a way as to change the general conclusions we have reached so far and that are nicely summarized in Fig.~\ref{radiofig}: a DM interpretation of the Fermi data is in strong tension with existing antiproton and, in particular, radio data.  While the  expected flux in antiprotons is not very sensitive to the form of the DM \emph{profile}, for example, the radio constraints in principle are. However, the profile cannot be changed significantly because of the need to fit also the angular dependence of the Fermi data down to an angular resolution of about $0.1^\circ$ (which corresponds to a distance of $\sim10\,$pc from the galactic center; note that a cutoff at this scale would be completely ad hoc and not have any astrophysical motivation -- already the cutoff at 1pc included in Fig.~\ref{radiofig} is artificially large and only added for comparison). The only point left to consider, then, is the possibility of modifying the \emph{annihilation channel}. For kinematic reasons, however, the annihilation to Higgs or gauge bosons is not possible; annihilation to other light quarks produces essentially the same spectrum in $\bar p$, $e^\pm$ and $\gamma$ as for $b\bar b$, and gamma rays radiated from light lepton final states would have a very different energy distribution compared to what would be required in order to reproduce the Fermi data. This leaves only one potentially interesting channel, i.e.~annihilation into $\tau^\pm$, but one may easily check that also in this case the resulting spectrum in gamma rays is too hard to fit the data in a satisfying way.

To summarize, the recent antiproton data very nicely confirm earlier expectations about the flux of secondaries and thereby provide interesting constraints on exotic contributions like from DM annihilation (a fact that has been already observed, at higher energies, in the earlier released $\bar p/p$ data \cite{Donato:2008jk}); especially at low DM masses these constraints on the annihilation cross section are quite relevant as they are on the order of what is expected for thermally produced WIMPs. Radio data are usually even more constraining. Taken together, these constraints are in strong tension with the possibility that the gamma rays observed by Fermi could be due to DM annihilation in the galactic center. The radio bounds are, in fact, so stringent that they seem to seriously obstacle even the discrimination of a potential sub-dominant DM contribution to the signal.

\section*{Acknowledgements}
I would like to thank Dan Hooper for a very helpful discussion and Mark Pearce for letting me use the preliminary PAMELA data presented in \cite{pam}. As an Emmy Noether fellow, I gratefully acknowledge support from the German Research Foundation (DFG) under grant BR 3954/1-1.


\end{document}